\documentstyle[epsf,aps,preprint]{revtex}
\begin{document}
\draft
\preprint{IMSc-96/05/13;~hep-th/9607006}
\title{Eikonal Approach to Planck Scale Physics}
\author{Saurya Das
\footnote{E-Mail:~saurya@imsc.ernet.in}}
\address{The Institute of Mathematical Sciences, \\ CIT Campus,
Madras - 600 113,  India.}
\maketitle
\begin{abstract}
We consider gravitational scattering of point particles with Planckian
centre-of-mass energy and fixed low momentum transfers in the framework of 
general relativity and dilaton gravity. The geometry around the 
particles are modelled by arbitrary black hole metrics of general 
relativity to calculate the scattering amplitudes. However, for dilaton 
gravity, this modelling can be done {\it only} by extremal black hole metrics. 
This is consistent with the conjecture that extremal black holes are 
elementary particles. 

\end{abstract}
%\pacs{04.60. -m, 04.62. +v, 11.80.Fv}
\newpage

\section{Introduction}
It is well known that the quantum effects of gravity come into play at 
the Planck energy scale (which is about $M_{pl}\sim 10^{19}~GeV$). The 
space-time 
curvatures become too large for classical general relativity to hold good 
at this scale. There has been various attempts to understand these effects 
and to formulate a quantum theory of gravitation. Notable among them are 
string theory and the Ashtekar formalism. However, till date, there is  
no fully satisfactory renormalisable theory of quantum gravity. We 
address the issue of Planck scale physics in the context of particle 
scattering via gravitational interaction at very high energies. The kinematics of 
particle scattering can be expressed by the two independent Mandelstam 
variables $s$, and
$t$, which are Lorentz scalars. They are respectively the squares of the 
centre-of-mass energy and the momentum transfer in the scattering process. 
Newton's constant $G$ being a dimensional constant and equal to 
$M_{pl}^{-2}$,  
the Planck scale can arise in two ways, either when $Gs \sim 1$ or 
when $Gt \sim 1$. 
Thus, the most general quantum 
gravitational scenario involves both $s$ and $t$ approaching the Planck 
scale. 

The eikonal approximation, on the other hand, is characterised by 
scattering at high $s$ and fixed low $t$. Physically, this signifies 
scattering of particles at very high velocities (and kinetic energies) 
and at large impact parameters, such that the interaction is weak and the 
particles deviate slightly form their initial trajectories. In other 
words, they scatter almost in the forward direction. 
We would restrict our analyses 
to this approximation and try to extract whatever information is 
available about the Planck scale effects as reflected in the scattering 
amplitudes. The motivation to study this 
kinematical regime is, as we shall see, that the scattering amplitudes in 
this approximation can be exactly calculated and expressed in a closed 
form. These of course become significant only at Planckian 
centre-of-mass energies and we can obtain some quantitative results 
about quantum gravity in this kinematical domain. 

Without loss of generality, two-particle scattering processes 
are considered. 
An inertial frame is chosen, in which one of the particles move at 
almost the speed of light and the other is relatively slow. Then
one of these point particles is modelled as the source of 
an appropriate metric of general relativity. For example, neutral particles 
are modelled by Schwarzschild metric and electrically or magnetically charged 
particles are modelled by Reissner-Nordstr\"om metric. Finally, 
the quantum mechanical wave-function of the other particle in the 
background of this space-time is analysed to deduce the corresponding 
scattering amplitude. The effect of electromagnetism is also studied when 
the particles also carry electric and/or magnetic charges.

Next, we replace these black hole metrics by their counterparts in 
dilaton gravity, i.e. those that arise in low energy string theory. Here, 
we see that the above mentioned modelling cannot be done by generic black 
holes as was the case for general relativity. If we consider charged 
particles, for example, the modelling can be successfully done only by 
extremal black holes to be able to calculate the scattering amplitudes. 
This supports the conjecture that extremal black holes can indeed be {\it 
identified} with elementary particles. 

\section{Eikonal Scattering in General Relativity}
\label{gtr}

We begin by considering the scattering of neutral point particles. The 
space time around any such particle, when it is static,  is obtained by 
solving 
Einstein's equations and given by the well known Schwarzschild metric:
\begin{equation}
ds^2~=~\left(1 - \frac{2GM}{r} \right) dt^2 - \left( 1 - \frac{2GM}{r} 
\right)^{-1} dr^2 - r^2 \left( d\theta^2 + \sin^2\theta d\phi^2 \right)~.
\end{equation}
To obtain the corresponding space-time when the particle is moving at a 
very high velocity, say along the positive $z$-axis, we perform a Lorentz 
transformation on the metric tensor components with the velocity 
parameter $\beta$ according to:
$$ t'~=~\gamma \left( t - \beta z\right) ~,$$
$$ z'~=~\gamma \left( z - \beta t \right)~,$$
where $\gamma = 1/ \sqrt {1- \beta^2}$. Simultaneously, the mass $M$ is 
parametrised as 
$$M~=~\frac{P_0}{\gamma}~.$$
This is to ensure that the energy of the boosted particle remains finite 
and equals $P_0$. The metric tensor components transform as a symmetric
second rank tensor. 
Dropping the primes and taking the limit $\beta \rightarrow 1$, the final 
form of the metric is:
\begin{equation}
ds^2~=~dx^-\left[dx^+-\frac{2GP_0}{|x^-|}dx^-\right]-dx_{\perp}^2~~,
\label{boo}
\end{equation}
where $x^{\pm} \equiv t \pm z$, the light cone coordinates. The 
above geometry was first obtained in \cite{aich} and then in 
\cite{thdr}. Defining the new coordinate differentials $d{\tilde 
x}^\mu$ as $$d{\tilde x}^+~=~dx^+ - \frac{2GP_0}{|x^-|} dx^-~, \\
d{\tilde x}^-~=~dx^- ~, \\
d{\tilde x}_\perp ~=~ dx_{\perp}~,$$
Eq.(\ref{boo}) can be re-written as,
\begin{equation}
ds^2~=~d{\tilde x^-}d{\tilde x^+} - d{\tilde x}_\perp^2~.
\end{equation}
The above form of the infinitesimal line element seems to indicate that 
we have simply arrived at flat space-time by a coordinate re-definition. 
However, writing the finite forms of these re-definitions, we get:
\begin{eqnarray}
{\tilde x^+}~&=&~x^+ - 2GP_0 \theta \left( x^- \right) \ln x_\perp~,  
\label{shift}\\ 
{\tilde x^-}~&=&~x^- ~,\\ 
{\tilde x_\perp}~&=&~x_\perp~.
\end{eqnarray}

Note that the transformations are continuous everywhere except at 
$x^-=0$, which is the trajectory of the boosted particle, where there is a 
step function discontinuity in the coordinate $x^+$. Calculation of the 
Riemann-Christoffel curvature 
tensor reveals that they are Dirac-delta functions (derivatives of 
the $\theta$ functions), which are non-vanishing only at $x^-=0$. 
Thus, all space-time curvatures are localised on the  
two-dimensional transverse plane, perpendicular to the trajectory of the 
boosted particle and travelling along with it. We call this infinite 
plane the {\it shock-wave}. The space-time in front of and behind the 
shock-wave is Minkowskian. It is analogous to the case of a boosted 
charged particle, where the electromagnetic fields tend to become 
concentrated along the direction perpendicular to the particle 
trajectory, and in the limit $\beta \rightarrow 1$, they are completely 
localised on the plane fronted 
(electromagnetic) shock-wave. Note that the coordinate 
$x^-$ is however, continuous at all points and serves as a bonafide 
affine parameter for any test particle in the above background. 
The classical geometry is depicted in Fig(1), which can be thought of as 
two Minkowski spaces, pertaining to $x^-<0$ and $x^->0$ respectively 
and glued along the null plane $x^-=0$, which is the trajectory of the 
boosted particle and the shock-wave. 

\begin{center}
\leavevmode
\epsfxsize= 10cm
\epsfysize= 20cm
\epsfbox {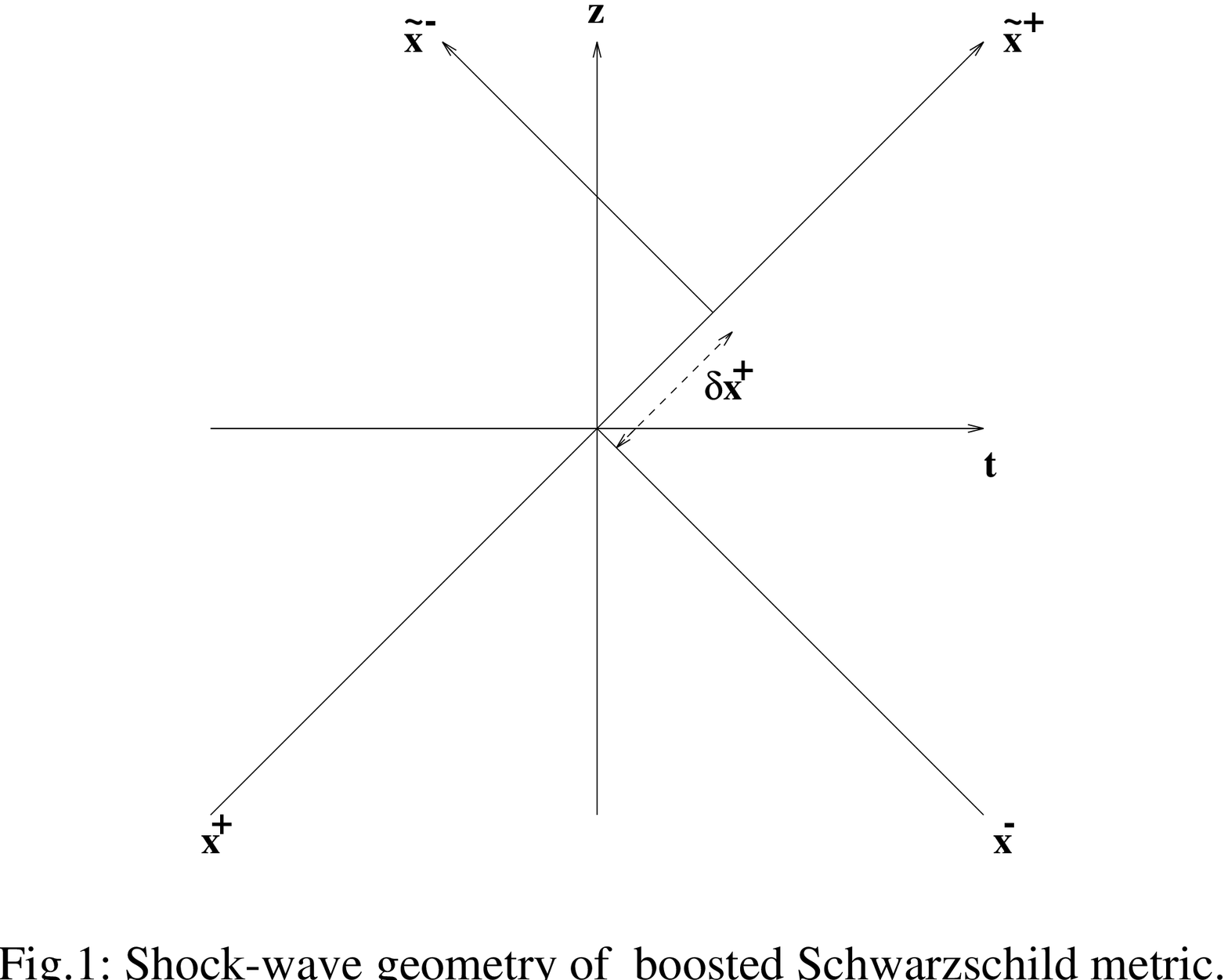}
\end{center} 

Having found the geometry of the luminal particle, now we concentrate on 
the other particle, assumed to be relatively slow. It serves as the 
test-particle in the above background. Before the shock wave comes and 
hits this particle, it is free from any interactions, and its 
wave-function is given by
\begin{equation}
\psi_<~=~e^{ip \cdot x}~=~e^{i\left[p_+x^+ + p_-x^- + \vec p_\perp 
\cdot \vec x_\perp \right]}~~,~~~~~x^-<0~~. 
\end{equation}
The moment when it is hit by the shock-wave, the $x^+$ coordinate 
undergoes a discrete shift given by Eq.(\ref{shift}) and the wave 
function picks up a space-time dependent phase factor. Simplifying, 
we get the final wave function to be:
\begin{equation}
\psi_>~=~e^{-iGs \ln x_\perp^2}~\psi_<~.
\end{equation}
Here we have used the identity $2p_-p_0 = s$. 
To calculate the scattering 
amplitude from this wave function, we expand it in terms of a complete 
set of momentum eigenstates and perform an inverse Fourier transform to 
obtain the expansion coefficients. The latter can be identified with the 
scattering amplitude modulo kinematical factors. The detailed calculation 
is done in ref. \cite{thf}, and the final result is:
\begin{equation}
f(s,t) = \frac{Gs}{t} \frac{ \Gamma(1-iGs)}{\Gamma (1+iGs)} \left( 
\frac{-1}{t} \right) ^{-iGs}~. \label{eik}
\end{equation}
Note that the above expression is simply the Rutherford Scattering 
amplitude with the gravitational coupling constant $-Gs$ replacing its 
electromagnetic counterpart $\alpha$. As advertised, it captures the 
gravitational interactions between point particles at the Planck scale 
and is insignificant for sub-Planckian energies, when $Gs \ll 1$. 

%\begin{center}
%\leavevmode
%\epsfxsize= 5in
%\epsfbox {ia1.eps}
%\end{center} 

Without going into the details, which the reader will find in \cite{dm3}, 
we summarise in brief the situation when electromagnetic interactions are 
included in the scattering process. That is, the scattering particles 
also carry electric or magnetic charges. If they carry electric charges 
$e$ and $e'$, then the scattering amplitude is modified to:
\begin{equation}
f(s,t) = \frac{Gs-ee'}{t} \frac{\Gamma\left(1-i(Gs-ee')\right)}{\Gamma 
\left(1+i(Gs-ee')\right)} \left(\frac{-1}{t}\right)^{-i(Gs-ee')}~. \label{em}
\end{equation}
The remarkable fact about this expression is that it can be obtained 
from the purely gravitational result in Eq.(\ref{eik}) simply by making the 
replacement $Gs \rightarrow Gs-ee'$. This means that the gravitational 
and electromagnetic coupling constants simply add up to give the 
effective coupling and there is no interference between them. This is 
quite unique and holds only in the eikonal approximation, because as we 
know, the two forces do affect each other in a non-trivial manner in 
generic cases. This decoupling is reminiscent of the Newtonian limit, 
where gravitation and electromagnetic interactions can be assumed not to 
affect each other. However, as far as the velocities and the energies of 
the particles are concerned, we are far removed from the Newtonian (low 
velocity) regime.

If, on the other hand, one of the particles carry an electric charge $e$, 
and the other a magnetic charge $g$, then the scattering amplitude is 
\cite{dm1}:
\begin{equation}
f(s,t)~=~ \left( \frac{n}{2} - iGs \right) \frac{ \Gamma
\left(\frac{n}{2} - iGs\right)}{\Gamma \left( \frac{n}{2} + iGs
\right)} \left( \frac{-1}{t}\right)^{1-iGs}
\label{eg}
\end{equation}
Here, the two couplings do not add up in a simple manner as in the 
previous case, but the same in not expected intuitively, because 
with magnetic monopoles, the interaction is no longer central in 
nature like gravitation or electromagnetism involving charges only. 
Comparing Eqs. (\ref{em}) and (\ref{eg}), we find that the
electromagnetic contribution to the former is insignificant (with $ee'$ 
typically of the order of the fine structure constant $1/137$), while for 
the latter, it is comparable to the gravitational contribution (both 
couplings being of the order of unity). In short, gravitation dominates 
overwhelmingly over electromagnetism at Planckian energies in the absence 
of magnetic monopoles. Introduction of the latter entails drastic changes 
in the result.

\section{Shock Waves in Dilaton Gravity}
\label{dg}

As noted earlier, the results in the previous section were derived in the 
framework of general relativity, and all the black hole metrics used to 
model the particles satisfy the Einstein's equations. Now, string theory, 
in the low energy limit provides us an alternative theory of gravitation, 
known as {\it dilaton gravity}, where in addition to the metric tensor, a 
scalar field called the {\bf dilaton} is also an 
independent degree of freedom. We will not go into the details as to how 
this theory emerges as a low energy {\it effective} theory from string 
theory. Instead, we will analyse the scattering situation envisaged in 
section \ref{gtr} using dilaton gravity. We would like to investigate 
whether the scattering amplitudes are modified in this framework, and 
whether the electromagnetic and gravitational decoupling still hold good.
We will see that modelling the scattering particles by dilatonic black 
holes poses some generic pathologies, which are removable only under 
certain specific conditions and when these are satisfied the decoupling 
exists just as in the case of general relativity. 
The counterpart of the Reissner-Nordstr\"om metric in dilaton gravity is 
given by the following expression (in the so-called `string metric')
\cite{gar}:
\begin{equation}
ds^2~=~\left(1 - \frac{\alpha}{Mr} \right)^{-1} \left[ \left(1- 
\frac{2GM}{r}\right) dt^2 - \left( 1- \frac{2GM}{r}\right)^{-1} dr^2 
\right] - r^2 d\Omega^2~,
\label{dgm}
\end{equation}
where $\alpha=Q^2 e^{2\phi_0}$, $Q$ being the electric charge and 
$\phi_0$ the asymptotic value of the dilaton field. Note that setting 
$Q=0$ reproduces the Schwarzschild metric, which means that the dilaton 
field has non-trivial effects only when we consider charged solutions. 
Like the Reissner-Nordstr\"om solution, the above metric has two horizons 
$r_+$ and $r_-$, given by:
$$r_+~=~2GM~,$$
$$r_-~=~\frac{\alpha}{M}~.$$
However, here a crucial difference is that the inner horizon $r_-$ is a 
space-time singularity, where the curvature tensor blows up. For black 
holes of large masses, this singularity is hidden behind the event 
horizon $r_+$ and there is no naked singularity, which however develops as the 
mass decreases. We will see that this singularity plays a crucial role in 
eikonal scattering. Analogous to the Schwarzschild or the 
Reissner-Nordstr\"om metric, we apply a Lorentz boost to the dilaton 
gravity metric along the positive $z$-axis and take the limit $\beta 
\rightarrow 1$. The resultant metric is of the form \cite{gar}:
\begin{equation}
ds^2~=~dx^- \left[ \frac{1 - \alpha/2P_0 |x^-|}{1 - \alpha/P_0|x^-|} \right]
\left[dx^+ - \frac{4GP_0/|x^-|}{1-\alpha/P_0|x^-|}~dx^- \right] - 
dx_\perp^2~.
\end{equation}
As before, to express this in a form representing Minkowski space, we 
define the new coordinates
\begin{eqnarray}
d{\tilde x}^-~&=&~dx^- \left[\frac{1-\alpha/2P_0|x^-|}{1-\alpha/P_0|x^-|}
 \right]\\
d{\tilde x}^+~&=&~dx^+ - \frac{4GP_0/|x^-|}{1-\alpha/P_0|x^-|}~dx^-~~.
\end{eqnarray}
Note that here both the coordinates $x^-$ and $x^+$ are discontinuous 
for non-vanishing $\alpha$. This is rather disturbing, because as we saw 
in the previous section, $x^-$ served as the continuous affine parameter 
for the test particle, and a discontinuity in it signals a breakdown in 
the description of the evolution of the particle in terms of geodesics. 
This is indeed confirmed by writing the classical geodesic equations of 
the test particle in the background of the boosted dilaton metric and 
trying to solve it perturbatively in powers of the mass $M$. The failure 
of the latter indicates that the null geodesics are incomplete in this 
case and the curvature singularity at $r^-=\alpha/M$ shows up as an 
extended naked singularity in the boosted limit and renders eikonal 
scattering impossible. The classical geometry in this case is shown in 
Fig.(2), where the rectangular shaded region denotes the finite region of 
singularity that originated from the singular horizon $r_-$. Thus, we see 
that modelling the particles by dilaton gravity metric gives rise to 
singularities which makes the subsequent calculation of scattering 
amplitudes impossible.

\begin{center}
\leavevmode
\epsfxsize= 10cm
\epsfbox {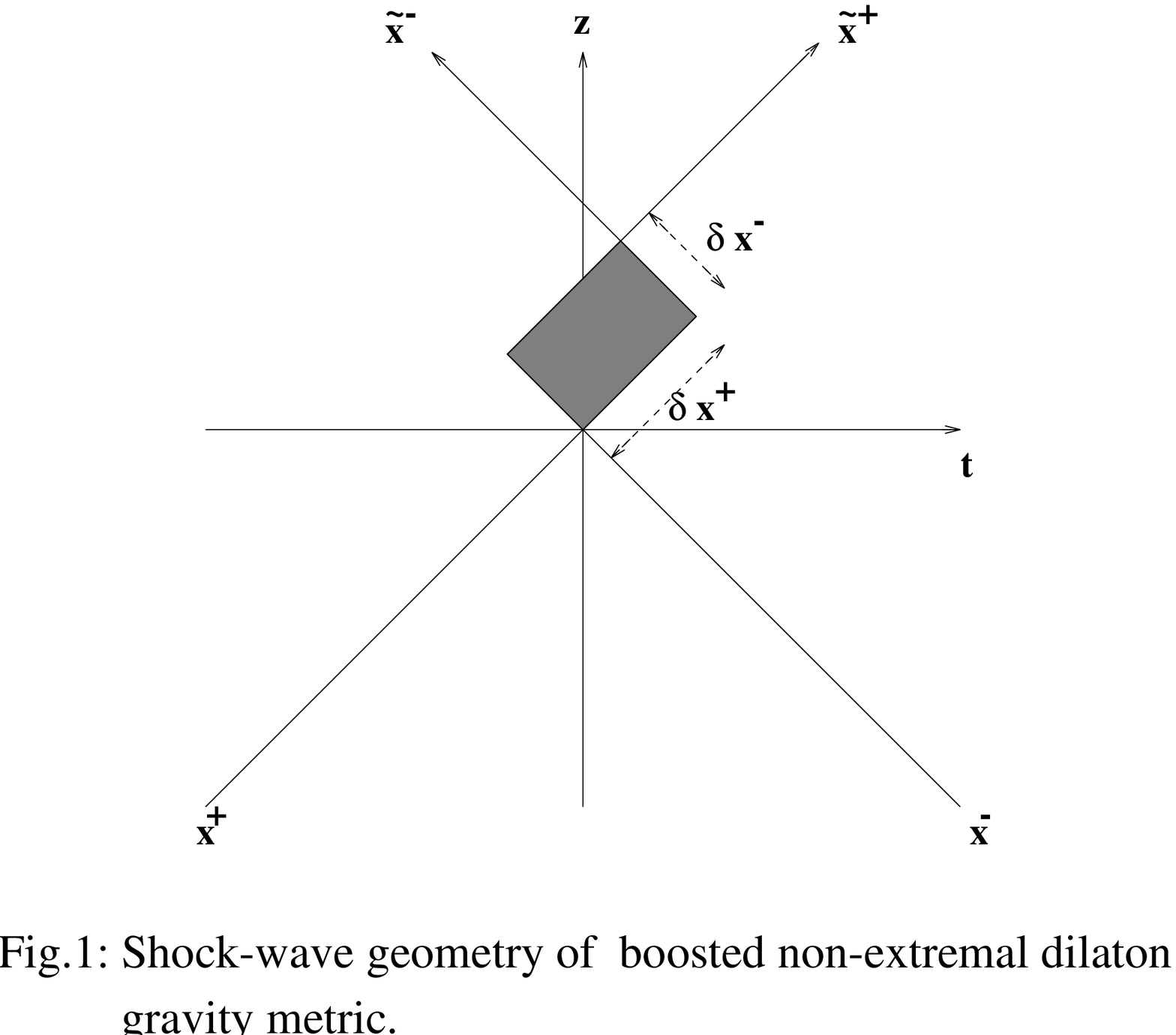}
\end{center} 
To circumvent this difficulty, we can try various possibilities. In
particular, we can examine the 
case by 
imposing the {\it extremal limit} $r_+=r_-$. Then, the metric assumes the form 
\begin{equation}
ds^2~=~dt^2 - \frac{dr^2}{\left(1 - \frac{2GM}{r}\right)^2} 
-r^2d\Omega^2 ~,
\end{equation}
which is perfectly regular everywhere. The curvature singularity has 
simply disappeared when the two horizons coincide ! This is precisely the 
motivation behind considering this limit. Note that the metric above is 
entirely distinct from the Schwarzschild metric. However, on performing the 
usual Lorentz boost on it along with the limit $\beta \rightarrow 1$, it 
can be easily verified that the boosted version coincides with the 
metric (\ref{boo}). This is a pleasant surprise, because now the test 
particle will `see' an identical background as in the case of 
Schwarzschild, and the corresponding scattering amplitude is simply the 
eikonal result (\ref{eik}). 

Thus it is clear that the curvature singularity shows up as an extended 
singular region and makes it impossible to calculate scattering 
amplitude. The extremal limit, on the other hand, plays a special role 
whose imposition reproduces the elegant eikonal results. In the next
section, we will probe into the details of the role of this singularity 
by invoking a different formalism.

\section{Scattering in Static Dilaton Gravity Background}
\label{ext}
In this section, we try to establish rigorously, the role of the 
naked singularity in dilaton gravity metric in eikonal scattering. Here 
we consider the scattering of the fast particle in the background of the 
other static particle. Note that the definitions of `source' and 
`test' has been reversed. The Klein-Gordon equation for the wave function 
$\phi$ of the fast particle is given by:
\begin{equation}
g_{\mu \nu}~D^\mu D^\nu \Phi~=~0 ~,
\end{equation}
where $D^\mu$ denotes the covariant derivative. The above equation can be 
simplified to
\begin{equation}
\frac{1}{{\sqrt -g}}\partial_\mu \left({\sqrt -g} g^{\mu \nu} \partial_\nu 
\Phi \right) ~=~0~.
\label{kg}
\end{equation}
We assume a solution of the form
\begin{equation}
\Phi~=~f(r)~Y_{lm}(\theta,\phi)~e^{iEt}~,
\end{equation}
where we have split the wave function into a radial part, the  
spherical harmonics and the exponential of asymptotically measured energy 
$E$. This decomposition results from the spherically symmetric and static 
nature of the dilaton gravity metric (\ref{dgm}). With this ansatz, the 
following radial equation is obtained from Eq.(\ref{kg})
\cite{dm5}:
\begin{equation}
r^2\Lambda \frac{d^2 f}{dr^2} + 
\frac{d\left(r^2\Lambda\right)}{dr}~\frac{df}{dr} - \left[ 
\frac{l(l+1)}{\Delta} - \frac{E^2r^2}{\Lambda} \right]~f~=~0~,
\label{rad}
\end{equation}
where we have defined the quantities $\Lambda \equiv 1 - 2GM/r$ and 
$\Delta \equiv 1- \alpha/Mr$. As expected, we recover the radial equation 
for Schwarzschild background for $\alpha = 0,\Delta=1$, and the subsequent 
scattering amplitude (\ref{eik}) \cite{dm4}. However, for generic values 
of $\Delta$, it can be shown that the above radial equation does not 
admit of a solution at all. This follows from an elementary theorem in 
ordinary differential equations since the coefficient of $f$ 
in eq.(\ref{rad}) suffers an infinite discontinuity at $\Delta=0$. 
Thus, as in the previous section, we conclude that the dilaton gravity 
metric (\ref{dgm}) cannot be used to successfully model the high energy 
particles. Moreover, we are now in a position to understand the physical 
origin of the pathology. The curvature singularity at $r_-=\alpha/M$ 
grows indefinitely large as we take the limit $M \rightarrow 0$ and 
eventually fills all space around the dilatonic particle. Thus, the other 
particle, at arbitrary 
impact parameter, is forced to hit this 
singularity and get trapped, signaling the breakdown of the scattering  
process. This was reflected in the non-existence of solutions of the 
classical geodesic equations in section \ref{dg}. Imposing the extremal 
limit, on the other hand, eliminated this singularity, and thus the 
scattering amplitude became calculable, which yielded the eikonal result.
This can be verified using the Klein-Gordon radial equation also. 
In the extremal limit, Eq.(\ref{rad}) assumes the form \cite{dm4}:
\begin{equation}
\frac{d^2f}{dr^2} + \frac{1}{r^2\Lambda} 
\frac{d\left(r^2\lambda\right)}{dr} \frac{df}{dr} - \frac{1}{\Lambda^2} 
\left[\frac{l(l+1)}{r^2} - E^2 \right]~f~=~0~.
\end{equation}
The above radial equation can be expanded in powers of $GM/r$ and this 
recovers the Schwarzschild radial equation to the lowest order. The 
scattering amplitude (\ref{eik}) follows immediately. This re-emphasises 
the importance of the extremal limit for Planckian scattering via dilaton 
gravity. 

\section{Perturbative approach}
\label{per}

In the previous two sections, we have explicitly used a classical solution
of the low energy string effective action to arrive at the scattering
amplitudes (in the extremal limit). However, the problem can be approached
without the help of such explicit solutions, at the level of the action
itself. Historically, the eikonal scattering amplitude was derived in
the context of quantum electrodynamics by
summing an infinite subset of Feynman diagrams, known as {\it ladder
diagrams}, with certain kinematical restrictions on the matter
propagators. It was shown that this infinite sum can be expressed in a
neat closed form \cite{abar}. A crucial assumption required to arrive at 
the eikonal result is the assumption that the scattering particles have 
well defined {\it classical} trajectories, which differ slightly from the 
free particle trajectories. The sum of ladder diagrams were seen to
converge for gravitational scattering as well in \cite{kabat} which 
reproduced the amplitude (\ref{eik}) exactly. For dilaton gravity, 
however, the assumption regarding classical trajectories is invalidated 
because, as we saw in the previous sections, an incoming particle is 
swallowed up by the expanding curvature singularity and there are no well 
defined scattering solutions. Thus, a priori, it seems impossible to 
construct and calculate ladder diagrams from dilaton gravity action given by:
\begin{equation} S~=~\int d^4x {\sqrt -g}e^{-2\phi} \left[
- \frac{R}{G} + F_{\mu \nu}F^{\mu \nu} + 2 \partial_\mu \phi \partial^\mu
\phi \right]~. \label{act} 
\end{equation} 
First, we simplify the action by
linearising the metric as well as the dilaton field, 
\begin{eqnarray}
g_{\mu \nu}~&=&~\eta_{\mu \nu} + h_{\mu \nu} ~, \\ \phi~&=&~\phi_0 + f~,
\label{lin} \end{eqnarray} 
where $\eta_{\mu \nu}$ is the flat Minkowskian
metric and $\phi_0$ is some constant. Retaining terms to leading order in
these quantum fluctuations, the action (\ref{act}) reduces to:
\begin{eqnarray}
S~&=&~\frac{e^{-2\phi_0}}{G} \int d^4x (1 -2f) \frac{1}{8} h_{\mu \nu} 
\left[ \eta^{\mu \lambda} \eta^{\nu \sigma} + \eta^{\mu \sigma}\eta^{\nu 
\lambda} - \eta^{\mu \nu} \eta^{\lambda \sigma} \right]\Box   h_{\lambda 
\sigma}  \nonumber \\
&-&e^{-2\phi_0} \int d^4x \left( 1 + \frac{1}{2}h_\alpha^{~\alpha} 
\right) (1 - 2f) \left[ -4 \partial_\mu f \partial^\mu f + F^2 + 
\partial_\mu \chi \partial^\mu \chi \right]~.
\end{eqnarray}
Here we have also included the action for the massless matter field $\chi$ 
representing the scattering particles. In addition to the graviton and matter
propagators and the matter-graviton interaction vertex already calculated 
in \cite{kabat}, now we have a dilaton propagator and a matter-dilaton 
interaction vertex. The factors associated with them can be read of fom 
the linearised action, and turns out to be $-i/{(p^2+m^2 -i\epsilon})$ and 
$-2p \cdot p'$ respectively, where $p$ and $p'$ are the momenta associated 
with the external matter lines. With these, the new infinite set of 
ladder diagrams with dilaton exchanges can be computed in a 
straightforward manner. The details of the calculation is done in 
\cite{dm5}, and the final result is:
\begin{equation}
i{\cal M}~=~\frac{ip_1^2p_2^2}{-t} 
\frac{\Gamma(1-ip_1^2p_2^2/Ep)}{\Gamma(1+ip_1^2p_2^2/Ep)} 
\left(\frac{4}{-t}\right)^{-i\frac{p_1^2 p_2^2}{Ep}}~.
\end{equation}
Now, if we make the momenta $p_1$ and $p_2$ on-shell and replace them by 
$m^2$ and eventually take the massless limit $m \rightarrow 0$, then we 
see that the dilaton amplitude vanishes identically and we are simply 
left with the gravitational result (\ref{eik})! 

At this point, we investigate the circumstances under which the action 
can be linearised, because without the latter, the eikonal sum can never 
be attempted. The metric $g_{\mu \nu}$ is linearised under the assumption 
that there are small graviton fluctuations over a Minkowskian 
background. As for the dilaton field, we can try to estimate its quantum 
fluctuation $f$ by looking at the classical solution for $\phi$ obtained 
by minimising the action (\ref{act}):
\begin{equation}
e^{2\phi}~=~e^{2\phi_0} \left( 1 - \frac{\alpha}{Mr} \right)~,
\end{equation}
which implies that the fluctuations over $\phi_0$ are of the order of
\begin{equation}
f \sim |\phi - \phi_0| \sim |\ln \left(  - \alpha/Mr \right)|~.
\end{equation}
Smallness of this requires $|\alpha/Mr| \rightarrow 0$, or in other words 
$alpha$ should scale at least as $M^2$ as we take $M \rightarrow 0$. But 
this is equivalent to the extremal limit, when $\alpha=2GM^2$ ! Thus, 
even in a solution-independent approach, where we look at the action and 
impose certain restrictions on it, the extremal limit seems to emerge in 
a natural way, if we want to obtain well defined scattering amplitudes. 

\section{Conclusion}
Our analysis of Planckian scattering in the light of dilaton gravity 
unambiguously point to the fact that there are important constraints to be 
satisfied while trying to model the scattering particles by a suitable 
metric. Namely, the extremality condition should be necessarily imposed 
on the parameters to obtain physically meaningful scattering amplitudes. 
Otherwise, the curvature singularity at a finite radius inevitably shows 
up as pathologies in the calculations. In the shock-wave approach, where 
we boosted the dilaton gravity metric, this was seen to give rise to the 
a discontinuous affine parameter. Next, while trying to solve the 
Klein-Gordon equation in the background of the above metric, we saw that 
there were no scattering solutions to the corresponding radial equation. 
Finally, in the approach of perturbation theory, we showed that the 
eikonal scattering amplitude can be reproduced if and only if the dilaton 
field was linearised and the lowest order terms in its quantum 
fluctuation is retained in the action. Once again, this linearisation is 
consistent with the extremality condition. There is yet another 
solution-independent way to arrive at identical conclusions, starting 
from the dilaton gravity action, where one uses the so called 
`Verlinde-scaling' to incorporate the eikonal kinematics. The interested 
reader may refer to \cite{dm5} for a detailed discussion of this 
approach. The important point to note is that no such restrictions were 
ever necessary in the general relativistic framework to calculate 
scattering amplitudes. Thus, it is perhaps correct to say that the theory 
of gravity that emerges from string theory incorporates certain 
problematic features, at least in the context of Planckian scattering. 
But the same theory contains the cure to this problem also, namely in the 
form of the extremal limit! The latter constraint, once imposed, removes the 
pathologies altogether and reproduces the finite amplitudes of general 
relativity. It is also curious to note the consistency of these results 
with the well known conjecture that extremal black holes are actually 
elementary particles \cite{sen}. Here, we are considering scattering 
of point particles, which can also be regarded as `elementary'. Thus, it 
seems logical in the spirit of the conjecture, to model them as extremal 
black holes.


\begin{references}
\bibitem{aich} P. Aichelburg and R. Sexl, Gen. Rel. Grav. {\bf 2} (1971)
303.
\bibitem{thdr} T. Dray  and G. 't Hooft, Nucl. Phys. {\bf B 253} (1985) 173.
\bibitem{thf} G. 't Hooft, Phys. Lett. {\bf B 198} (1987) 61; Nucl. Phys.
{\bf B 304} (1988) 867.
\bibitem{jac} R. Jackiw, D. Kabat and M. Ortiz, Phys. Lett. {\bf B
277} (1992) 148.
\bibitem{dm3} S. Das and P. Majumdar, Phys. Rev. {\bf
D 51} (1995) 5664.
\bibitem{dm1} S. Das and P. Majumdar, Phys. Rev. Lett. {\bf 72}
(1994) 2524.
\bibitem{gar} H. Garfinkle, G. Horowitz and A. Strominger, Phys. Rev.
{\bf D 43} (1991) 3140; {\it err.} Phys. Rev. {\bf D 45} (1992) 3888.
\bibitem{dm2} S. Das and P. Majumdar, Phys. Lett. {\bf B 348} (1995) 349.
\bibitem{dm5} S. Das and P. Majumdar, IMSc preprint 95/25,
hep-th/9512209 (1995).
\bibitem{dm4} S. Das and P. Majumdar, IMSc preprint 95/8,
hep-th/9504060 (1995).
\bibitem{ince} E. L. Ince, Ordinary Differential Equations (Dover,
{}~New
York, 1956), p. 73.
\bibitem{abar} H. D. I. Abarbanel and C. Itzykson, Phys. Rev. Letters
{\bf 23} (1969) 53;
 M. L\'evy and J. Sucher, Phys. Rev. {\bf 186} (1969)
1656.
\bibitem{kabat} D. Kabat and M. Ortiz, Nucl. Phys. {\bf B 388} (1992)
570.
\bibitem{ver} H. Verlinde and E. Verlinde, Nucl. Phys. {\bf
B 371} (1992) 246.
\bibitem{sen} A. Sen, Mod. Phys. Lett. {\bf A 10} (1995) 2081 and
references therein.

\end{references}
\end{document}